\documentclass{ws-rv9x6}
\usepackage{ws-rv-van}
\usepackage{bm}
\usepackage{graphics}         
\makeindex

\newcommand{\bq}{\begin{equation}}
\newcommand{\ee}{\end{equation}}
\newcommand{\fr}[2]{\frac{#1}{#2}}

\begin{document}

\chapter[Spin-Hall Effect in Chiral Electron Systems]{Spin-Hall
Effect in Chiral Electron Systems: from Semiconductor
Heterostructures to Topological Insulators\label{ch1}}

\author[F. Author and S. Author]{P.G.Silvestrov$^1$ and E.G.Mishchenko$^2$
}

\address{
 $^1$Theoretische
Physik III, Ruhr-Universit{\"a}t Bochum, 44780 Bochum, Germany \\
 $^2$Department of Physics, University of Utah, Salt Lake
City, Utah 84112, USA }


\begin{abstract}
The phenomenon of mesoscopic Spin-Hall effect reveals in a
nonequilibrium spin accumulation (driven by electric current) at
the edges of a ballistic conductor or, more generally, in the
regions with varying electron density. In this paper we review our
recent results on spin accumulation in ballistic two-dimensional
semiconductor heterostructures with Rashba/Dresselhaus spin orbit
interactions, and extend the method developed previously to
predict the existince of spin-Hall effect on the surface of
three-dimensional topological insulators. The major difference of
the new Spin-Hall effect is its magnitude, which is predicted to
be much stronger than in semiconductor heterostructures.
This happens because in semiconductors the spin accumulation
appears due to a small spin-orbit interaction, while the
spin-orbit constitutes a leading term in the Hamiltonian of
topological insulator.

\end{abstract}

\body

\section{Chiral electron systems}

Chiral systems feature electron states whose quantum properties
(e.g. spin) depend on the direction of propagation. One example of
such chiral states arises from spin-orbit coupling that originates
from relativistic corrections to the dispersion law of band
electrons. In particular, for a typical two-dimensional electron
gas (2DEG) the intrinsic asymmetry of a confining quantum well
geometry is accompanied by a strong perpendicular ($z$) electric
field that leads to spin-orbit coupling of the Rashba
type~\cite{BR},
 \begin{equation}\label{rashba}
H_R=\lambda {\bf z} \cdot ({\bm \sigma} \times {\bf
p}) =\lambda(\sigma_xp_y-\sigma_yp_x).
 \end{equation}
As a result, the  spin degeneracy is lifted via {\it effective
momentum-dependent} Zeeman field, ${\bf h}_{\bf p}=\lambda
(-p_y,p_x)$. The electron spin eigenstate is thus determined by its
momentum, ${\bf p}$.

Another important type of spin-orbit coupling occurs in 2DEG
formed by semiconductors with broken inversion symmetry, e.g.\
GaAs, InAs. While only third order in momentum in the bulk GaAs,
this coupling, known as Dresselhaus interaction \cite{Dress}, is
``upgraded'' in two dimensions to the linear order  by virtue of
transverse momentum quantization and becomes,
\begin{equation}
\label{linearDress}
 H_{D}= \lambda_D (\sigma_xp_x-\sigma_yp_y).
 \end{equation}
Chiralities acquired from different spin-orbit couplings
(\ref{rashba}) and (\ref{linearDress}) are {\it opposite}, in a
sense that the electron wave function acquires opposite Berry
phases, $\pm \pi$, upon (adiabatic) completion of a loop in the
momentum space (enclosing the degeneracy point ${\vec p}=0$),
depending on whether $\alpha>\beta$ or vice versa.

Dirac fermions in graphene represent another realization of the
chiral system \cite{ACN}. Due to hexagonal symmetry of underlying
two-dimensional honeycomb atom arrangement the low-energy electron
excitations are combined into two Dirac cones (valleys), $K$ and
$\widetilde{K}$, within the first Brillouin zone. The corresponding
effective Hamiltonians are
\begin{equation}
H_K=v(\tau_xp_x+\tau_yp_y), ~~~~ H_{\widetilde K}
=v(\tau_xp_x-\tau_yp_y).
\end{equation}
Here ${\bm \tau}$ stands for the {\it pseudospin} operator that
acts in the sublattice space.

It is interesting to note the analogy between {\it total} graphene
Hamiltonian and a semiconductor with {\it equal} Rashba and
Dresselhaus couplings ($\alpha=\beta$) in the  case of a
conventional 2DEG. The latter case features spin eigenstates that
are momentum-independent \cite{SEL} and, thus, somewhat trivial.
However, since the two cones in graphene are {\it separated} in
momentum space, the chiral physics can still be observed. (Yet some
phenomena are trivially absent, e.g. pseudospin-Hall effect, cf.
last section). It is important to emphasize that the chirality in
graphene has nothing to do with spin-orbital coupling (which is
rather weak in carbon allotropes) and is a consequence of the
crystal geometry.

Another notable example of a chiral electron system is a
topological insulator. The latter is different from the usual band
insulator in that its valence and conduction bands originate not
from different atomic orbitals but from {\it the same
spin-orbit-split} orbital. In 2D HgTe quantum wells this gives
rise to topologically protected edge states leading to a recently
predicted\cite{BHZ} and discovered\cite{Kon} quantum spin-Hall
effect. Yet even more intriguing twist has recently been added by
a discovery of two-dimensional states on the surface of 3D
topological insulators\cite{Hsi,Hsi1} Bi$_{0.9}$Sb$_{0.1}$,
Bi$_{2}$Sb$_3$, and Bi$_{2}$Te$_3$. The fundamental difference
from graphene is that the number of pockets of the Fermi surface
within the Brillouin zone is {\it odd}, with the complementary
species residing on the opposite surface of a sample. Thus, these
states are not simply separated in the momentum space (like in
graphene), but also separated in {\it real} space. This removes
the above mentioned ``trivialization'' that is present in
graphene. In particular, spin-Hall effect can occur in topological
insulators. As confirmed by the first-principle band structure
calculations\cite{HJZ} spin structure of these states is indeed
chiral. In particular, the low-energy Hamiltonian of
Bi$_{2}$Te$_3$ can be deduced from symmetry considerations to be
of the form\cite{Fu},
 \begin{equation}
\label{linearD}
 H_{TI} =v(\sigma_x p_y-\sigma_y p_x) +\frac{p^2}{2m^*} + {\alpha} \sigma_z
 p_x(p_x^2+3p_y^2).
 \end{equation}
For an ungated and undoped Bi$_{2}$Te$_3$ the last  term is
generally of the same order of magnitude as the first two. Still, as
a starting point it is useful to neglect the effects of anisotropy.

\section{Spin-Hall effect}

Spin-Hall effect \cite{DP} is the name given to a class of phenomena
that exhibit boundary (surface, edge) spin polarization when
electric current flows through a system with significant spin-orbit
interaction. It has been  observed in both 3D \cite{exp1,Kato,exp3}
and 2D systems \cite{exp2}. It is customary to distinguish two
mechanisms that could lead to this effect. The {\it extrinsic}
mechanism is the dominant one in 3D semiconductors and originates
from scattering off impurities~\cite{Hirsch99,Z,ERH,DS}. Presence of
impurities is unavoidable in high carrier density 3D semiconductors
that rely on doping. Quite contrary, {\it intrinsic}
mechanism~\cite{MNZ,Sinova} originates from spin-orbit-split
band-structure, which induces spin precession when electric current
is driven through the system. This mechanism can in principle be
realized in coexistence with ballistic transport in 2D electron
systems systems. Indeed, by placing dopant far enough from
GaAs/AlGaAs interface one can reduce effects of disorder scattering.

In the present paper we concentrate on the intrinsic spin-Hall
mechanism. It is known, however, in two-dimensional electron
systems with spin-orbit coupling linear in momentum (typical for
$n$-doped heterostructures) {\it any} scattering that leads to a
stationary electric current via deceleration of electrons by
impurities, phonons, etc., will negate the precession due to
external electric field and result in the exact cancellation
\cite{IBM,MSH,Kh,RS,D} of the bulk spin-current in a dc
case\footnote{This exact cancellation does not occur in 3D- or 2D
hole-systems that feature non-linear spin-orbit couplings.}.

There are several ways to avoid such cancellation, in particular,
to use ac currents with frequencies exceeding the inverse spin
relaxation time \cite{MSH,Ora2005}. The second possibility and the
one of interest to us here, is make a system sufficiently small
and clean (ballistic) so that the electron mean free time exceeds
the time of flight across the systems. The corresponding scenario
is known as the \emph{mesoscopic spin-Hall effect} \cite{nik}.
While initial theories of spin-Hall effect in infinite systems had
addressed the auxiliary quantity of spin current (for a review see
Refs.~\cite{reviews,ERH1}), in a finite geometry it is both easier
and more relevant to calculate spin polarization density,  which
is an experimentally measurable quantity~\cite{nonlocal}. Such
edge polarization was considered by numerical methods in several
earlier publications~\cite{NWS,nik,nik3,usaj,SG} as well as both
analytically and numerically in our previous papers
\cite{ZSM,SZM}.

A crucial note is due. The edge spin polarization in ballistic
systems appears not as a result of electric field-driven
acceleration of electrons and associated with it precession of
spins.  Indeed, electric field in a ballistic system is absent as
the electric potential drop occurs over a contact region with the
leads rather than over a bulk of a ballistic conductor.
Nevertheless, spin precession does occur. It originates from
accelerated electron motion in the boundary potential that
provides lateral confinement. Bias applied to the contacts ensures
that populations of left- and right-moving states are different
and the net spin precession appears. Naturally, it is proportional
to the applied bias $V$. The net spin accumulation near an edge of
a 2DEG is {\it independent} of the shape of the boundary
potential,
\begin{equation} \label{result} \int_{-\infty}^\infty s_z(x) dx = -
\frac{\lambda^2-\lambda_D^2 }{12\pi^2 v^3_F}eV,
 \end{equation}
where $v_F$ is the bulk value of the Fermi
velocity\footnote{Strictly speaking, the integration in
Eq.~(\ref{result}) goes from a pint far outside the 2DEG ($-\infty$)
to some point deep inside 2DEG ($+\infty$), but still far away from
the edges. Integration across the whole conductor would give zero,
reflecting the fact that spin accumulation at the opposite edges has
opposite signs.}. Spin accumulation appears in the second order in
spin-orbit interaction and vanishes for equal Rashba and Dresselhaus
coupling strengths.

Position-resolved spin polarization $s_z(x)$ can be found
analytically in two important situations. First is the case of a
smooth confining potential $U(x)$ and the second is an infinite
hard-wall boundary. We begin with analyzing semiclassical electron
motion in smooth potentials in Sec.~\ref{Semiclassics}. In
Sec.~\ref{Hard Wall} we derive the result (\ref{result}) for the
net spin accumulation and illustrate it using the case of
hard-wall boundary. In Sec.~\ref{Soft Wall} we present a method of
kinetic equation that allows to find local spin-polarization for
smooth boundary potentials. In Sec.~\ref{Turning} singular
dynamics near classical turning points is discussed. Finally, in
Sec.~\ref{Topological} we use methods developed in the preceding
chapters to describe spin-Hall effect in topological insulators.

Many of the results presented in this paper were published in the
journal articles Refs.~[\refcite{ZSM,SZM,SM}]. However, the
derivation of the spin accumulation via the Kinetic equation in
Sec.~\ref{Soft Wall} and prediction of the nonequilibrium spin
accumulation in topological insulators in Sec.~\ref{Topological}
are presented here for the first time

\section{Semiclassical electron motion}\label{Semiclassics}

Consider gated 2DEG with Rashba spin-orbit
interaction\footnote{Similar methods may be used to describe more
complicated interactions, like Dresselhaus~\cite{Dress}, or
combined Rashba and Dresselhaus interactions.}, described by the
Hamiltonian
 \bq\label{HamBook}
H=\fr{p^2}{2m} +\lambda
(p_y\sigma_x-p_x\sigma_y)+\fr{m\lambda^2}{2}+U(x,y).
 \ee
Potential $U(x,y)$ is created by the external gates, or the edge
potential ensuring the in-plane confinement of 2DEG. Possible effect
of disorder on $U(x,y)$ are going to be neglected.

Classical electron dynamics described by the short wave length
limit of the Hamiltonian~(\ref{HamBook}) reveals a number of very
unusual features. As a first step, in this section we demonstrate
how the trajectories-based approach allows to describe propagation
of fully (in-plane) polarized electric currents through mesoscopic
constrictions.

Construction of semiclassical solutions of the Schr\"odinger
equation with the Hamiltonian~(\ref{HamBook}) follows the
reasoning of the conventional WKB
approach~\cite{Lj,Bolte98,Plet03,Culcer04,SM}, which is valid for
a smooth potential, $\hbar |\nabla U|\ll \min(p^3/m, p^2\lambda)$.

Without the external potential $U$, the electron spectrum consists
of the two subbands, $E_\pm(p_x,p_y)=(p\pm m\lambda)^2/2m$. The
subbands meet at only one point, $p=0$, and the spin in each
subband is always aligned with one of the in-plane directions
perpendicular to the momentum $\vec{p}$. The semiclassical
electron dynamics~\cite{Lj} naturally captures the essential
features of this translationally invariant limit. The classical
motion in each subband is determined by the equations of motion
which follow from the effective Hamiltonian:
 \bq\label{Hameff}
H_{\rm eff}=\fr{(p\pm m\lambda)^2}{2m} +U(x,y).
 \ee
Despite the fact that spin does not appear in this equation, one
can easily construct semiclassical wave functions, which have spin
pointed within the $xy$~plane perpendicular to the momentum:
 \bq\label{wf}
 \psi=u e^{iS/\hbar} , \ \
 u=\sqrt{\fr{\rho}{ {2p}}}
 \left(\begin{array}{cc} \sqrt{p_y+ip_x}\\
 \pm\sqrt{p_y-ip_x} \end{array}\right).
 \ee
Here the action $S$ is related to the momentum by $\vec{p}=\nabla
S$, and $\rho=u^\dag u$ is the classical density for a family of
classical trajectories corresponding to a given energy $E$. The
action $S$ obeys the classical Hamilton-Jacobi equation,
 \bq
\fr{(|\nabla S|\pm m\lambda)^2}{2m}+U(x,y)=E.
 \ee

During its motion, an electron described by Eq.~(\ref{wf}) changes
the momentum $p$ but always remains in the same spin-subband. To
change the subband the electron trajectory should pass through the
degeneracy point where both components of momentum vanish
simultaneously, $\vec{p}=0$, which is generically impossible.
Moreover, with the proper use of potential barriers, one may
realize a situation where electrons {\it of only one subband} are
transmitted and the others are totally reflected. This leads to
strong in-plane polarization of the transmitted electron flow.

To take into account the out-of-plain spin precession one has to
go beyond the approximation Eq.~(\ref{wf}), as was done in
Ref.~[\refcite{SM}]. Instead of doing so we will describe in
Sec.~\ref{Hard Wall} a method allowing to find easily the
expectation value of $\sigma_z$ for potential depending only on
one coordinate $U=U(x)$ (boundary potential).

\begin{figure}
\centerline{\psfig{file=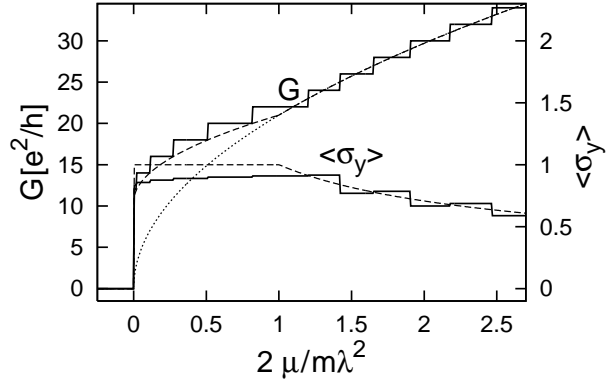,width=8.cm}}
 \caption{ Conductance (in units of $e^2/h$), and spin polarization of the
current vs gate voltage (in units of $m\lambda^2/2$). Dashed lines
show the smoothed curves (\ref{Landauer}),(\ref{Spin}), solid
lines show the quantized values for $m\lambda L/\hbar=10.5\pi$.
Dotted line shows the conductance without spin-orbit interaction.
} \label{fig1.1}
\end{figure}

\subsection{ Sharvin conductance}\label{Sharvin}

To give an example of a spin-polarized current let us consider
transmission through a barrier, $U(x)$, varying along the
direction of a current propagation. We assume periodic boundary
conditions in the perpendicular direction ($y+L\equiv y$), thus
$p_y$ is an integral of motion\footnote{Spin polarized currents on
a cylinder with $x$-dependent spin-orbit interaction were
considered recently in Ref.~[\refcite{Ora}]}. For a smooth
potential $U(x)$ the conduction channels may either be perfectly
transmitting or completely closed. The conserved transverse
momentum takes the quantized values, $p^{n}_{y}=2\pi\hbar n/L$.
Consider the functions
 \bq
E^{n}_\pm(p_x)=\fr{(p^n\pm m\lambda)^2}{2m}  , \ \
p^n=\sqrt{p_x^2+{p^{n}_{y}}^2}
 \ee
For $n\neq 0$ the function $E^{n}_\pm(p_x)$ splits into two
distinct branches. At any point $x$ the equation
 \bq\label{channel}
E^{n}_\pm(p_x) = E_F -U(x)
 \ee
yields solutions $p_x^L$ and $p_x^R$, corresponding to left- and
right-moving electrons. Application of a small bias implies, e.g.,
the excess of right movers over left movers far to the left from
the barrier. Particles are transmitted freely above the barrier if
Eq.~(\ref{channel}) has a solution, $p_x^R$, for any $x$. Let
$\mu=E_F-U_{\max}$ be the difference between the Fermi energy and
the maximum of the potential. The $n$th channel in the upper
branch opens when
 \bq\label{muplus}
\mu=\left({2\pi\hbar |n|}+{m\lambda L} \right)^2/2mL^2.
 \ee
For the lower branch $E_-^{n}(p_x)$ Eq.~(\ref{channel}) has four
solutions (two for right and two for left movers) for $|n|<
m\lambda L/2\pi\hbar$ and $x$ close to the top of the barrier.
However, far from the barrier (where the excess of right-movers is
created) there are still only two crossings described by
Eq.~(\ref{channel}), one for right and one for left movers. As a
result, all the extra electrons injected at $x=-\infty$ follow the
evolution of a solution of Eq.~(\ref{channel}) with the largest
positive $p_x$. For all $|n|< m\lambda L/2\pi\hbar$ such a
solution does exist for any positive $\mu$. Thus, at $\mu=0$ as
many as $n_0=m\lambda L/\pi\hbar$ channels {\it open up
simultaneously}. The channels with higher values $|n|>m\lambda
L/2\pi\hbar$ in the lower subband $E_-^{n}$ open when
 \bq\label{muminus}
\mu=\left({2\pi\hbar |n|}-{m\lambda L} \right)^2/2mL^2.
 \ee
According to the Landauer formula, ballistic conductance is given
by the total number of open channels multiplied by the conductance
quantum $G_0=e^2/h$
 \bq\label{Landauer}
G=G_0\fr{ L}{\pi\hbar} \left\{\begin{array}{cl}
 \sqrt{{2\mu m}}+m\lambda,~~
& 0<\mu <m\lambda^2/2 \\
 2 \sqrt{{2m\mu}}, &
\mu >m\lambda^2/2. \end{array}\right.
 \ee
This dependence $G(\mu)$ is shown in Fig.~\ref{fig1.1}. The
striking evidence of the presence of spin-orbit interaction is the
huge jump of the conductance at the pinch-off point, as opposed to
the conventional square-root increase in the absence of spin-orbit
coupling. This jump is a consequence of the ``Mexican-hat'' shape
of the spectrum $E_-(p_x,p_y)$. Accuracy of Eqs.~(\ref{muplus})
and~(\ref{muminus}) is sufficient to resolve the steps in the
conductance due to the discrete values of $|n|=0,1,2,...,$
(conductance quantization), as shown in Fig.~\ref{fig1.1}. The
steps in $G(\mu)$ are abrupt in the limit $dU/dx\rightarrow 0$.

Close to the pinch-off, at $\mu\lesssim m\lambda^2$, the conserved
$p_y$ component of the electronic momentum varies for different
transmitted channels within the range $|p_y|\lesssim m\lambda$.
Therefore, far from the barrier, where the Fermi momentum is large
$p_F \gg m\lambda$,  we have $p_x\gg p_y$ and transmitted
electrons propagate in a very narrow angle interval
$|\theta|<\sqrt{m\lambda^2/2E_F}\ll 1$. Since the electron spin is
perpendicular to its momentum, we conclude that the current due to
electrons from each of the subbands is almost fully polarized. The
total polarization of the transmitted current is given by the
difference of two currents
 \bq\label{Spin}
\langle{\sigma_y}\rangle=
\fr{{\langle\psi^\dagger{\sigma_y}v_x\psi
\rangle}}{{\langle\psi^\dagger v_x\psi\rangle }}= \min
(1,\sqrt{{m\lambda^2}/{2\mu}}),
 \ee
which is also depicted in Fig.~\ref{fig1.1}. This current
polarization may also be viewed as a creation of in-plain
nonequilibrium spin density, maximal on the barrier.

Derivation of Eqs.~(\ref{Landauer}) and~(\ref{Spin}) was greatly
simplified because of the periodic boundaries. Our next example
demonstrates semiclassical treatment of realistic boundary
conditions.

\subsection{ Quantum Point Contact}\label{QPC}

\begin{figure}
\centerline{\psfig{file=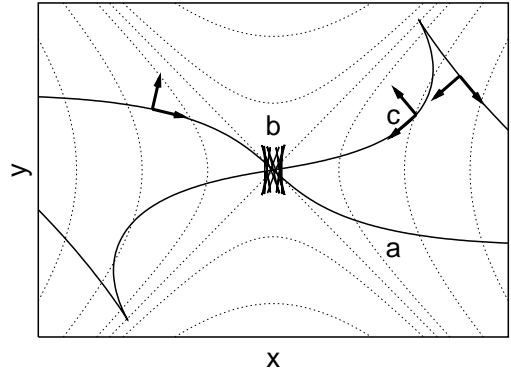,width=7.3cm}}
 \caption{ Three kinds of trajectories in the point contact. $a$,
transmitted trajectory whose momentum is always collinear with the
velocity. $b$, trajectory bouncing inside the QPC. This trajectory
is periodic in the linearized approximation described in the text,
while the exact calculation for finite amplitude shows its slow
drift. $c$, transmitted trajectory whose momentum inside the
contact is opposite to the velocity. Electrons flow from left to
right. Arrows show momentum and spin orientations. Few
equipotential lines are also shown. } \label{fig1.2}
\end{figure}

Let us consider probably the most experimentally relevant example
of a quantum point contact, described by the potential
 \bq\label{VQPC}
U(x,y)= -\fr{m\Omega^2x^2}{2} +\fr{m\omega^2 y^2}{2}.
 \ee
Even in this simple model the electron flow in the presence of
spin-orbit interaction acquires a number of interesting and
peculiar features. This become clear already from the
figure~\ref{fig1.2}, where we show three types of trajectories in
such potential. Each kind of trajectories is necessary for
calculation of conductance.

Classical equations of motion follow in the usual manner from the
effective Hamiltonian (\ref{Hameff}): $\dot{\vec{r}}=\partial
H_{\rm eff}/\partial \vec{p}$, $\dot{\vec{p}}=-\partial H_{\rm
eff}/\partial \vec{r}$. We consider quantum point contact~(QPC)
close to the  opening with only the lower $E_-$ subband
contributing to the conductance. A crucial property of the
Hamiltonian $H_{\rm eff}$, Eq.~(\ref{Hameff}), is the existence of
a circle of minima of the kinetic energy at $|p|=m\lambda$.
Expanding around a point on this circle,
$p_{x_0}=m\lambda\cos\alpha$, $p_{y_0}=m\lambda\sin\alpha$, one
readily finds the equations of motion for ${\cal P}
=p_x\cos\alpha+p_y\sin\alpha- m\lambda\ll m\lambda$,
 \bq\label{motimotion}
 \ddot{\cal P}+(-\Omega^2\cos\alpha^2+\omega^2\sin\alpha^2){\cal P}=0
 \ , \
 \dot{\alpha}=0.
 \ee
The trajectory is found from the relations, $\dot{x}={\cal
P}\cos\alpha/m \ , \ \dot{y}={\cal P}\sin\alpha/m$. We observe
from Eq.~(\ref{motimotion}) that only the trajectories within the
angle
 \bq\label{tanalpha}
\tan|\alpha|< \tan\alpha_0={\Omega}/{\omega}
 \ee
are transmitted through QPC. Trajectories with larger angles are
trapped (oscillate) within the point contact. Examples of both
types of trajectories are presented in Fig.~\ref{fig1.2}.
Quantization of trapped trajectories would give rise to a set of
(extremely) narrow resonances in the conductance, specific for
spin-orbit interaction. Below we consider only the smoothed
conductance.

To calculate the current $J$ through QPC one has to integrate over
the phase space of the states which are transmitted from left to
right,
 \bq
J=\int\limits dy\int\limits ev_x\fr{d^2p}{(2\pi\hbar)^2}=GV,
 \ee
and have the energy within the interval $\mu-eV/2< E_{-}<
\mu+eV/2$, with ${V}$ standing for the applied voltage. In this
section we define $\mu$ as the difference between the Fermi energy
and the value of the potential at the saddle point
$\mu=E_F-U(0,0)$. The integral is most simply evaluated at $x=0$
(with the velocity given by $v_x={\cal P}\cos\alpha /m$). The
allowed absolute values of the momentum are
 \bq
2\mu-{e{V}}-{m\omega^2y^2} < {{\cal P}^2}/{m}<
2\mu+{e{V}}-{m\omega^2 y^2}.
 \ee
The angle interval of transmitting trajectories consists of two
domains: $|\alpha|<\alpha_0,~{\cal P}>0$, and
$|\alpha-\pi|<\alpha_0,~ {\cal P}<0$. The appearance of the latter
range of integration is highly non-trivial. A simple reasoning
shows that the particles with the velocity antiparallel to the
momentum ($v_x>0$, $p_x<0$) should not contribute to the
conduction in the case of a transition through a one-dimensional
barrier $U=U(x)$, see Eq.~(\ref{Landauer}). Despite corresponding
to the right-moving electrons, these states {\it do not originate}
in the left lead. Indeed, they exist only in the vicinity of
$x=0$, but disappear as $x\rightarrow -\infty$ and, thus, cannot
be populated by the excess electrons (except due to the tunneling
transitions which are irrelevant in the semiclassical regime).
Such trajectories, however, {\it do exist} in QPC,
Eq.~(\ref{VQPC}), as demonstrated in Fig.~\ref{fig1.2}. After
passing through QPC the trajectory bounces at the wall reversing
its velocity. This kind of classical turning points, where both
components of the velocity vanish simultaneously, are specific for
the effective Hamiltonian~(\ref{Hameff}). The existence of
transmitting trajectories with $|\alpha-\pi|<\alpha_0,~ \varrho<0$
results in the doubling of the conductance. Simple calculation
yields
 \bq\label{conduc}
G=G_0\fr{4m\lambda\sin\alpha_0} {\pi\hbar\omega}
\sqrt{\fr{2\mu}{m}}.
 \ee
The presence of a threshold angle $\alpha_0$, as well as the
square-root dependence of $G(\mu)$, are in a sharp contrast to the
well-known result $G=G_0\mu/\pi\hbar\omega$, in the absence of
spin-orbit interaction.

Since Eq.~(\ref{motimotion}) describes only the linearized
electron dynamics, Eq.~(\ref{conduc}) is formally valid if $\mu\ll
m\lambda^2$.\footnote{Still the number of open channels should be
large for semiclassics.} Nevertheless, the current remains totally
polarized for $0<\mu < m\lambda^2/2$ [similar to Eq.~(\ref{Spin})]
 \bq\label{SpinQPC}
\langle{\sigma_y}\rangle=
\fr{{\langle\psi^\dagger{\sigma_y}v_x\psi
\rangle}}{{\langle\psi^\dagger v_x\psi\rangle }}= 1.
 \ee
With increasing the chemical potential, $\mu > m\lambda^2/2$,
transmission via the upper subband  $E_{+}$ kicks in and the
degree of polarization gradually decreases, like it happened in
Eq.~(\ref{Spin}).

In InAs-based heterostructures, typical value of spin-orbit
coupling \cite{Grun} is $\lambda\hbar =2\times 10^{-11}eVm$.
Characteristic spin-orbit length $l_R=\hbar/m^*\lambda = 100 $ nm
and energy $m^*\lambda^2/2= 0.1$ meV. In order to have strongly
spin-polarizing QPC, the latter should support many transmitting
channels at chemical potential $\mu\sim m^*\lambda^2/2 \gg
\hbar\omega$. This condition can, equivalently, be written in
terms of the width of the point contact $\Delta y$, see
Eq.~(\ref{VQPC}), as $\Delta y \gg l_R$.

\section{Edge spin accumulation. Hard wall}\label{Hard Wall}

\begin{figure}
\centerline{\psfig{file=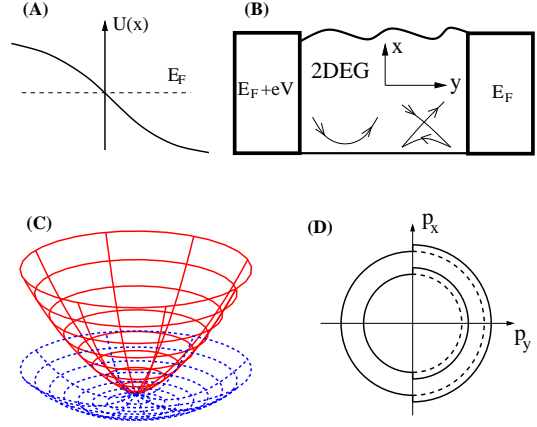,width=7.cm}}
 \caption{ Profile of a boundary potential $U(x)$. B) Geometry of the system:
two-dimensional electron gas ($x>0$) is attached to two ideal
reflectionless metallic leads filled by equilibrium electrons up
to different chemical potentials. C)
Spin-orbit-split subbands structure. D) Difference in population
of left- and right-moving electrons due to the applied bias $eV$.
} \label{fig1.3}
\end{figure}

Consider a semi-infinite ballistic 2DEG described by the
Hamiltonian Eq.~(\ref{HamBook}), where potential depending only on
one coordinate, $U\equiv U(x)$, ensures boundary confinement (see
Fig.~\ref{fig1.3}). The system is attached to two ideal
reflectionless leads injecting equilibrium electrons into 2DEG.
The chemical potentials of the leads are shifted by the applied
voltage, $eV$. The current flow along the $y$-direction in case of
spin-orbit interaction Eq.~(\ref{HamBook}) results in an edge spin
accumulation $s_z(x)$. In this section we first derive an exact
formula for the total amount of spin accumulated at the edge,
$\int s_z dx$. Then we consider in more details the case of hard
wall potential, $U(x<0)=\infty$, $U(x>0)=0$, where exact results
for $s_z(x)$ are available\footnote{Nonequilibrium edge spin
accumulation at the hard wall boundary of disordered conductor was
considered in Ref.~[\refcite{Sonin}]}.

\subsection{Net spin accumulation}\label{Net_Spin}

For the potential independent on $y$ ($U=U(x)$) the corresponding
momentum component, $k_y$, is an integral of motion. It is
convenient to use the Fourier representation along the $y$-axis
for the electron operators, $\hat \psi({\bf r})=\sum_{k_y}\hat
c_{k_y}(x)e^{ik_yy}$. We employ here the second quantization
formalism. One can derive the equation of motion for the
expectation value of the electron spin operator, ${\bf
s}(k_y,x)=\frac{1}{2} \langle \hat c^\dagger_{k_y}(x) \hat {
\sigma} \hat c_{k_y}(x)\rangle$, which can be readily written in
the form,
\begin{equation}
\label{conserv}
\partial_t s_y(k_y,x)=- \partial_x
J^y_x(k_y,x)-2\lambda k_y s_z(k_y,x).
\end{equation}
Here $J_x^y$ stands for the conventional operator of spin-current,
i.e.,
$$ J^y_x(k_y,x)=\frac{i}{4m}\langle \nabla_x \hat c^\dagger_{k_y}
\hat\sigma_y \hat c_{k_y} -  \hat c^\dagger_{k_y} \hat \sigma_y
\nabla_x \hat c_{k_y}\rangle -\frac{\lambda}{2}\langle \hat
c^\dagger_{k_y} \hat c_{k_y} \rangle.
$$
In a steady state the lhs of Eq.~(\ref{conserv}) vanishes.
Integrating Eq.~(\ref{conserv}) over the $x$-direction, we obtain
for the net spin polarization,
\begin{equation}
\label{polar} \int_{-\infty}^\infty s_z(x) dx =
-\frac{1}{2\lambda} \sum_{k_y} \frac{1}{k_y} J^y_x(k_y,\infty).
\end{equation}
It is straightforward to calculate the value of the
($k_y$-resolved) spin current $J^y_x(k_y,\infty)$ inside the bulk
of a 2D system:
\begin{equation}
\label{spin_current} J^y_x(k_y,\infty)=
-\frac{1}{2}\sum_{\beta=\pm 1} \sum_{k_x}
\left(\lambda+\frac{\beta k_x^2}{mk} \right) n_\beta(k_x,k_y),
\end{equation}
where $n_\beta(k_x,k_y)$ stands for the population of different
momentum states in the subband $\beta$. Only ``uncompensated''
states contribute to the non-equilibrium spin polarization given
by Eqs.~(\ref{polar}-\ref{spin_current}); these states describe
electrons that originate in the left lead ($k_y>0$) and belong to
the energy interval near the Fermi energy,
$E_F<(k+\beta\lambda)^2/2m < E_F+eV$. The integral (\ref{polar})
diverges logarithmically at $k_y \to 0$. Assuming the same
infrared cut-off in both subbands, $\widetilde k$, we observe that
the diverging $\ln{\widetilde k}$-contributions in the two
subbands cancel each other, yielding in the linear (in $V$)
response,
 \begin{equation}
\label{s_density} \int_{-\infty}^\infty s_z dx
=\frac{eV}{2\lambda(2\pi)^2} \left(\frac{2\lambda}{v_F} -
\ln{\frac{v_F+\lambda}{v_F-\lambda}}\right)
 \end{equation}
where $v_F=\sqrt{2E_F/m}$ is the Fermi velocity. Expanding this
{\it general} result to the lowest non-vanishing order in
$\lambda/v_F$ we recover the net boundary polarization, Eq.
(\ref{result}).

\subsection{Evanescent modes}

We are now going to consider the edge spin density in the case of
sharp (hard wall) edge potential. Since the sharp edge does not
impose any finite length scale, the question arise, what would be
the width of the edge spin distribution? Obvious candidate for
that comes from the evanescent modes~\cite{usaj}, whose wave
function do have an explicitly decaying component ($\sim
\exp(-m\sqrt{v_F \lambda}x)$).

The reflection at the sharp boundary mixes the two bulk subbands.
Evanescent contributions in the upper subband appear when the
reflecting states from the lower subband belong to the domain,
$m(v_F-\lambda)<k_y<m(v_F+\lambda)$. Repeating the calculations
leading to Eq.~(\ref{s_density}) but now for the evanescent domain
only, we obtain,
 \begin{equation}
\label{s_density_ev} \int_{-\infty}^\infty s_z^{\text{ev}} dx
=\frac{eV}{2\lambda(2\pi)^2} \left(2\sqrt{\frac{\lambda}{v_F}} -
\ln{\frac{1+\sqrt{\lambda/v_F}}{1-\sqrt{\lambda/v_F}}}\right).
 \end{equation}
Remarkably, the net evanescent contribution turns out to be much
larger than the full result Eq.~(\ref{s_density}). This means that
this contribution is {\it largely} cancelled by the contribution
from the normal domain $k_y<m(v_F-\lambda)$. Similar cancellation
of all smooth long wavelength contributions takes place for the
local spin density, as we discuss in the next section.

\begin{figure}
\centerline{\psfig{file=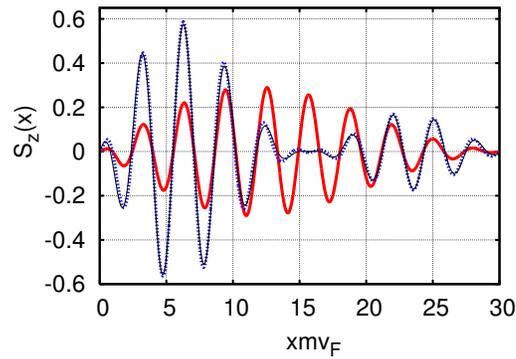,width=5.cm,angle=-90}}
 \caption{ Dependence of the local spin polarization,
in units of $eVm/8\pi^2$, on the distance to the boundary for
different values of spin-orbit coupling constant. Solid (red)
line: $\lambda/v_F=0.1$, dotted (blue) line: $\lambda/v_F=0.2$,
solid (black) line utilizes the approximate formula
(\ref{localapprox}) for $\lambda/v_F=0.2$. } \label{fig1.4}
\end{figure}

\subsection{Hard wall, results}

\begin{figure}
\centerline{\psfig{file=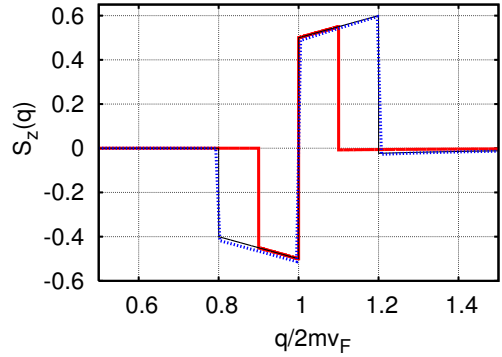,width=5.0cm,angle=-90}}
 \caption{ Spectral distribution (\ref{spectr})
of spin density in units of $ eV/4\pi v_{F}$ for different values
of spin-orbit coupling constant, $\lambda/v_F=0.1$, and
$\lambda/v_F=0.2$.
} \label{fig1.5}
\end{figure}

Exact lengthy explicit expression for $s_z(x)$ in the case of
sharp boundary potential, Fig.~\ref{fig1.4}, was given in
Ref.~[\refcite{ZSM}]. Here we show only the spectral density
 \bq
s_z(q)=2\int_{0}^\infty dx s_z(x) \sin{qx},
 \ee
which is given by a simple piecewise continuous algebraic function
defined in four domains. The surprising feature of the spectral
distribution is its vanishing, $s_z(q)= 0$, in the whole
long-wavelength domain, $0<q<2k^+$. In particular, this shows the
exact cancelation between normal and evanescent modes. For larger
values of $q$ we obtain to the leading order in $\lambda$,
 \begin{equation}\label{spectr}
s_z(q)=\frac{eV q}{16\pi m v_F^2} \left\{
\begin{array}{cl}  0, & q< 2m(v_F-\lambda),\\ -1, & 2m(v_F-\lambda) <q<2mv_F,\\ 1,&
2mv_F<q<2m(v_F+\lambda),
\\-2/(q\xi)^{4}, & 2m(v_F+\lambda) <q.
\end{array} \right.
 \end{equation}
The plot of the spectral distribution is illustrated in
Fig.~\ref{fig1.5}. Remarkably, the net spin polarization (given by
$\pi^{-1}\int dq s_z(q)/ q$) comes from the large-$q$ tail
($\propto q^{-3}$) in the spectral density $s_z(q)$.

The {\it approximate} spin density may be written in a simple form
($\hbar=1$),
 \begin{equation}
 \label{localapprox}
s_z(x)\approx \frac{eV}{2\pi^2v_F x}\cos{(2mv_Fx)}
\sin^2{(m\lambda x)}.
 \end{equation}
It is remarkable that the spin-orbit coupling constant enters via
the period of beating only.

As is evident from Eq.~(\ref{spectr}) one should speak about the
spin accumulation at the hard wall with certain caution. The spin
density in this case comes from quickly oscillating functions and
the notion of spin accumulation should be understood in the same
mathematical sense as the finite value of the integral
 \bq
\int_0^\infty dx \sin{x} =\lim_{\eta \to 0} \int_0^\infty dx~
e^{-\eta x}\sin{x}= 1.
 \ee

\section{Smooth edge: Kinetic equation}\label{Soft Wall}


The second ballistic spin-Hall problem that allows analytical
solution involves a smooth boundary potential, the exact condition
to be presented below. Away from classical turning points (see the
next section) spin accumulation can be obtained with the help of
the matrix kinetic equation. In addition this method allows to
find a complete electron distribution along the way. In the
semiclassical approximation the $2\times2$ matrix $\hat{f}_{\bf
p}(x)$ generalizes the usual distribution function. The equation
for the matrix distribution function in the presence of a smooth
confining potential $U(x)$ and in the absence of disorder takes
the form\cite{MH},
 \begin{equation}
\label{kinetic} \frac{p_x}{m} \partial_x \hat{f}_{\bf
p}-\lambda\left\{   \hat s_y,\partial_x \hat{f}_{{\bf p}}
\right\}+2i\lambda ~[p_y\hat s_x-p_x\hat s_y,\hat{f}_{{\bf p}} ]=
\frac{\partial \hat{f}_{{\bf p}}}{\partial{ p}_x}~\partial_x U(x).
 \end{equation}
To the zeroth order in the potential gradient $\partial_x U$ the
solution is trivial and reduces to the equilibrium Fermi-Dirac
distributions for the two spin-split subbands ($\beta =\pm 1$),
$f_{\beta}(x)=n_F(\frac{1}{2m}(p+\beta \lambda)^2+U(x)-\mu)$, with
the local value of the potential $U(x)$  determining the elevation
of the bottom of the subbands. To this order the electron spin
remains adiabatically {\it within} the plane of 2DEG during the
course of electron motion in the potential $U(x)$. The out-of-plane
spin component arises from the non-adiabatic corrections that are
first order in $\partial_x U$. The solution of kinetic equation
(\ref{kinetic}) is rather straightforward and yields,
 \begin{eqnarray}\label{denmatr}
\hat{f}_{{\bf p}}=\frac{1}{2}\sum_\beta [1+2\beta (n_y\hat
s_x-n_x\hat s_y)]f_{\beta}+ \frac{\lambda n_y \hat
s_z}{2p^2}\partial_x U \sum_\beta\frac{\partial (f_\beta
/\lambda)}{\partial \lambda},
 \end{eqnarray}
where ${\vec n}$ is the direction of the electron momentum. The
local value of spin-Hall density $s_z =\text{Tr}~\hat s_z \sum_{\bf
p} \hat{f}_{{\bf p}}$ is obtained by integrating Eq.~(\ref{denmatr})
over excess electron states that originate in the left lead, namely
over those with $v_y>0$ and the energy within the interval
$[E_F,E_F+eV]$,
 \begin{equation}
s_z(x) =  \frac{dU}{dx} \frac{m\lambda}{2(2\pi)^2} \frac{\partial
}{\partial \lambda} \frac{1}{\lambda}\int
\frac{dE}{\sqrt{2m[E-U(x)]}} \left(\frac{1}{p_+} -\frac{1}{p_-}
\right) [n_F (E-eV)-n_F(E)].
 \end{equation}
Here $p_{\pm}$ are the momenta corresponding to a given energy:
$p_{\pm}(x)=\sqrt{2m[E-U(x)]}\mp m\lambda$ when both subbands are
occupied, $\sqrt{2m[E-U(x)]}<m\lambda$; and $p_{\pm}=m\lambda \mp
\sqrt{2m[E-U(x)]}$ when the upper subband is empty,
$\sqrt{2m[E-U(x)]}>m\lambda$. Calculation of this integral in the
linear order in $eV$ yields
 \begin{equation}\label{local_spin}
\overline{s}_z (x) = - \frac{\lambda^2 eV}{(2\pi)^2m
(v_F^2-\lambda^2)v_F^3}\frac{dU}{dx}.
 \end{equation}

 The expression (\ref{local_spin}) has a number of interesting
 features.  The local spin polarization is proportional to the force
 exerted on electrons by the boundary. As long as the Fermi energy is well
 above the bottom of the bands, $v_F \gg \lambda$, spin accumulation
 is small and only second order in the spin-orbit coupling constant.
 When the bottom of the band is elevated high enough, $v_F <
 \lambda$, the local spin polarization increases dramatically. The
 vicinities of the two singularities at $v_F(x)=0$ (classical turning point) and
$v_F(x)=\lambda$ (degeneracy point) have to be studied by means
beyond kinetic equation (\ref{kinetic}).

In addition to a smooth classical spin distribution
(\ref{local_spin}) quantum wiggles in $s_z(x)$ are present whose
magnitude is not necessarily small compared to $\overline{s}_z
(x)$. A convenient quantity (especially for numerical
calculations) that averages out these wiggles is the "integrated
spin density"
 \bq
{\cal S}(x)=\sum_j\int_{-\infty}^x s_z (x')dx'.
 \ee
Integrating Eq. (\ref{local_spin}) over $x$ yields smooth part of
the integrated spin density
 \begin{equation} \label{s_density_smooth}
\overline {\cal{S}}(x) =\frac{eV}{2\lambda(2\pi)^2}
\left(\frac{2\lambda}{v_F(x)} -
\ln{\frac{v_F(x)+\lambda}{|v_F(x)-\lambda|}}\right).
 \end{equation}
This formula presents a generalization of Eq.~(\ref{s_density})
for the case of a smooth variation of the confining potential.

Note that the net spin polarization across the edge is again
independent of the shape of the boundary potential, $
{\cal{S}}(\infty) = - {\lambda^2 eV}/{12\pi^2 v^3_F}$, and is
expressed via the bulk value of the Fermi velocity $v_F(\infty)$.
Here by $x=\infty$ we assume a point deep inside the 2DEG but yet
far from its opposite edge. The latter has spin accumulation of the
same absolute value and opposite sign.

\section{Smooth edge: singular spin dynamics near turning
points}\label{Turning}

\begin{figure}
\centerline{\psfig{file=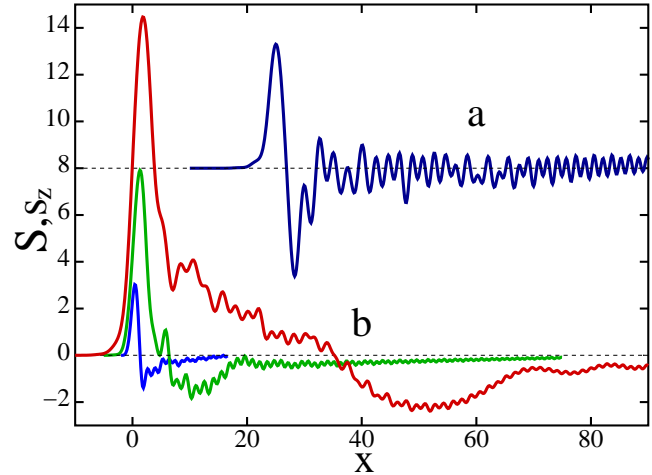,width=8.5cm}}
 \caption{a). Spin density for a force strength $F=0.01
m^2\lambda^3/\hbar$ ($U=-Fx$). [The curve offset both vertically
and horizontally.] All coordinates are measured in units of
$\hbar/m\lambda$. A smooth component of the density is hardly
visible because of oscillating contributions. b). The smooth
component is recovered in the integrated spin ${\cal
S}(x)=\int_{-\infty}^x s_z dx'$. Blue, green and red curves show
the integrated spin for the force strength $F=\alpha
m^2\lambda^3/\hbar$ with $\alpha = 0.25, 0.05, 0.01$,
respectively. In all these cases we see three regions with
different spin behavior. First, the spin density is the largest in
narrow outer ($x\approx 0$) strip along the edge. This spin is
compensated (and overcompensated) by the wide strip of negative
smooth spin density. Finally, in the third strip (at $x>50$ for
the red curve $\alpha =0.01$) the smooth component of the density
changes sign to positive again. The width of all three strips and
the amount of accumulated spin, which in each strip is much larger
than in Eq.~(\ref{result}), increase [formally unlimited] with
decreasing slope of the boundary potential. To obtain the values
of the spin density and accumulated spin one need to multiply the
numbers in the figure by $eVm/8\hbar\pi^2$ and $eV/8\lambda\pi^2$
respectively.} \label{fig1.6}
\end{figure}

Semiclassical Eqs.~(\ref{local_spin},\ref{s_density_smooth}) do
not offer the important information about the edge spin. First,
these equations predict a singular spin density at $v_F(x)=0$ and
$v_F(x)=\lambda$. In addition,
Eqs.~(\ref{local_spin},\ref{s_density_smooth}) ignore any
interference effects, which may be important for realistic
boundary potentials. Both these problems may be addressed
analytically, as it was done in Ref.~[\refcite{SZM}]. In order to
have more pedagogical discussion here we concentrate mostly on
numerical results. Let us approximate the boundary by a linear
potential
 \bq
U(x)=-Fx,
 \ee
with the constant force $F$.

Figure~\ref{fig1.6} shows the "raw" numerical data for $s_z(x)$
and ${\cal S}(x)$. The smoothness of the boundary implies that
$F\ll m^2\lambda^3/\hbar$. We see from Fig.~\ref{fig1.6}.a, how
the rapid quantum oscillations make it hard to observe the mean
value $\overline{s}_z(x)$ Eq.~(\ref{local_spin}) even for
$F\hbar/m^2\lambda^3 = 0.01$. The smooth component is recovered in
the integrated spin on Fig.~\ref{fig1.6}.b even for relatively
steep boundary $F\hbar/m^2\lambda^3 = 0.25$.

\begin{figure}
\centerline{\psfig{file=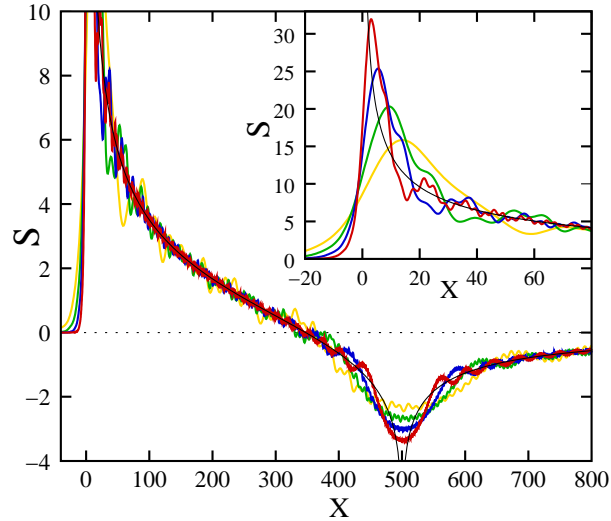,width=8.cm}}
 \caption{ Integrated spin density  ${\cal S}(x)=\int_{-\infty}^x
s_z dx$ for the potential $U(x)=-\alpha m^2\lambda^3 x/\hbar$ in
units of $eV/8\lambda\pi^2$. The curves for $\alpha = 8, 4, 2, 1
\times 10^{-3}$ are shown in yellow, green, blue and red
respectively. The horizontal coordinate is scaled differently for
different curves, as $x$ is measured in units of $10^3\alpha
\times \hbar/m\lambda$. Narrow black lines stand for the
semiclassical prediction, Eq.~(\ref{s_density}). The logarithimc
behavior, $\sim \log \alpha$, of the dip at at
$U(x)=-m\lambda^2/2$ ($x=500$) is clearly seen. Inset magnifies
the region near the edge of 2DEG ($x\approx 0$). } \label{fig1.7}
\end{figure}

Explicit comparison between numerics and analytical expression
Eq.~(\ref{s_density_smooth}) is made on Fig.~\ref{fig1.7}, where
we plot the rescaleed numerical results for different values of
the slope of the boundary potential.

According to Eq.~(\ref{s_density_smooth}) we find two regions of
different smooth spin behavior. First, within the strip where
$0<v_F(x)<\lambda$ spin density is negative (which is seen in a
downward slope of the integrated density ${\cal{S}}(x)$ in
Fig.~\ref{fig1.7}). Farther away, $s_z(x)$ changes sign for
$v_F(x)>\lambda$, where both $s_z(x)$ and ${\cal{S}}(x)$  decrease
gradually with increasing~$x$.

The most interesting is the behavior of spin  at the borders of
these regions, $v_F=0$ and $v_F=\lambda$. At $v_F(x)=\lambda$ the
accumulated spin $\overline {\cal{S}}(x)$
Eq.~(\ref{s_density_smooth}) diverges logarithmically. This
singularity originates from the accumulation of classical turning
points taking place when the conical crossing point in the
spectrum of the Hamiltonian (\ref{HamBook}), see
Fig.~\ref{fig1.3}C, passes through the Fermi energy. This
singularity is regularized as ${\cal{S}}\sim \log F$, according to
Fig.~\ref{fig1.7}.

Yet more peculiar is the behavior of both $s_z(x)$ and
${\cal{S}}(x)$ at the edge of 2DEG, near the point where
$v_F(x)=0$. The smooth part of the accumulated spin,
Eq.~(\ref{s_density_smooth}), has an infinite jump here (from
Eq.~(\ref{s_density_smooth}) it follows that $\overline
{\cal{S}}(+0)=\infty$, while obviously $\overline
{\cal{S}}(-0)=0$). Development of such jump with decreasing slope
of the potential is seen in the inset in Fig.~\ref{fig1.7}. The
jump in ${\cal{S}}(x)$ corresponds to the formation of a narrow
strip with extremely large values of spin $s_z>0$ along the
border. This behavior will now be analyzed in more detail.

\begin{figure}
\centerline{\psfig{file=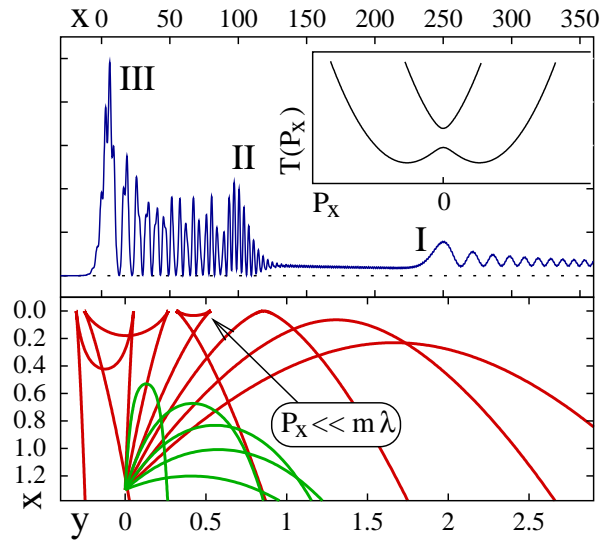,width=8.cm}}
 \caption{ {\it Bottom:} Family of classical trajectories at
$E=E_F$ for different values of $p_y$, and $m=\lambda=F$.
Trajectories for both lower (red) and upper (green) spin-orbit
split subbands are shown. An example of a trajectory contributing
to the peak in spin density at $x\approx 0$, Eq.~(\ref{stop}), is
shown for $"p_x\ll m\lambda"$ (see the text). {\it Top:} The
electron density $\rho =\psi_1^\dagger \psi_1 + \psi_2^\dagger
\psi_2$ for  given longitudinal momentum and energy. Here
$\psi_{1,2}$ are two eigenfunctions
with $p_y=0.2m\lambda, E_F=0,
F=0.003m^2\lambda^3/\hbar$, $\rho$ in arbitrary units, $x$ in
units of $\hbar/m\lambda$. Three classical turning points can be
seen. The interference of incoming and reflected waves in the
upper subband causes smooth oscillations to the right of the inner
turning point I ($x>m\lambda^2/2F$). At the other turning points,
II and III, the two kinds of oscillations are seen. Slow
oscillations are caused by the interference of the incoming wave
and the wave reflected at the turning point. Fast oscillations
(wavelength $\sim\hbar/m\lambda$) are due to the interference of
distant (in time) segments of the same trajectory. {\it Inset:}
Kinetic energy (arbitrary units) $T_\pm(p_x)=(p\pm m\lambda)^2/2m$
for fixed $p_y=0.2 m\lambda$. } \label{fig1.8}
\end{figure}

Classical dynamics of electrons with Rashba spin-orbit interaction
is described by the effective Hamilton Eq.~(\ref{Hameff}). with
the boundary potential approximated by the linear function
$U=-Fx$. The family of classical trajectories generated by this
Hamiltonian, shown in Fig.~\ref{fig1.8}, demonstrate a number of
unusual features.

As seen from Fig.~\ref{fig1.8}, those electrons from the lower
subband that have $|p_y|<m\lambda$, pass {\it three} turning
points in the course of their motion in the $x$ direction,
corresponding to three solutions of the equation $\partial {
H}_{\rm eff}/\partial p_x=0$. Two of these turning points (those
with $p=m\lambda$) correspond to simultaneous vanishing of {\it
both} velocity components, $\vec{v}(x)=0$, the behavior
generically impossible in a 2DEG with the parabolic spectrum,
${H}=p^2/2m$.

Analytical treatment~\cite{SZM} of the expectation value of the
$z$-component of electron spin in the vicinity of turning point
$v_F(x)=0$ yields ($x\ll m\lambda^2/F$)
 \begin{equation}
\label{szAiry} s_z =\frac{3meV}{4\pi\hbar}
\frac{\partial}{\partial \widetilde x} \int\limits_0^{1} dz~
\text{Ai}^2(- \widetilde x/z),
 \end{equation}
where $\widetilde x=x(2Fm/\hbar^2)^{1/3}$. In the asymptotic
region $x\gg (2Fm/\hbar^2)^{-1/3}$ one can average over the
oscillations of the Airy function. This allows us to recover the
singular behavior of the smooth spin density (\ref{local_spin}):
$\langle s_z\rangle\sim x^{-3/2}$. The integral in
Eq.~(\ref{szAiry})  features a logarithmic singularity at $x= 0$.
With the logarithmic accuracy the (properly regularized) height of
the peak of spin density is~\cite{SZM}
 \begin{equation}
\label{stop}
s_z(0)=\frac{meV}{10\sqrt{3}\pi^2\hbar}\ln{\left(\frac{m^2\lambda^3}{\hbar
F} \right)}.
 \end{equation}
Striking feature of this result is that this maximal value is
virtually {\it independent} of the strength of spin-orbit coupling
or the shape of the boundary potential (up to a weak logarithmic
factor).

We thus conclude that the nonequilibrium spin-Hall spin
accumulation near a smooth boundary of 2DEG ballistic conductor
with spin-orbit interaction develops a narrow peak at the edge,
with the width $\sim (\hbar^2/m F)^{1/3}$ and height given by
Eq.~(\ref{stop}). It is followed by a slow non-monotonic decay, as
shown on Fig.~\ref{fig1.7}. This smooth tail of spin density
persists to much larger distances, $\gtrsim m\lambda^2/F$. The
amount of spin accumulated in the peak (found as a maximum of the
function ${\cal S}(x)=\int^x s_z dx$) equals ${\cal S}_{\rm max}
\approx 0.04 eV(m^2/\hbar F)^{1/3}>0.04 eV/\lambda$, where in the
last inequality we utilize the fact that $F< m^2\lambda^3/\hbar$.
We see that the spin accumulated at the edge described by a
semiclassical boundary potential is inversely proportional to the
strength of spin-orbit interaction and becomes progressively
larger for smoother slopes. This prediction can be used for
experimental observation of spin-Hall effect in realistic
two-dimensional electron systems.

\section{Spin-Hall effect on surfaces of topological
insulators}\label{Topological}

The method developed in Sections \ref{Hard Wall} and \ref{Soft Wall}
is fully applicable to other chiral systems mentioned in the
Introduction, as long as the corresponding terms in the Hamiltonians
are {\it linear} in electron momentum. This is a good approximation
in graphene and an acceptable one in Bi$_2$Te$_3$. In graphene,
however, chiral structure arises from {\it pseudospin} (in
sublattice space). There is nothing straightforward about
experimental detection of pseudospin in graphene. In addition, as
explained in the Introduction, the case of graphene is mapped onto
``Rashba-like'' chirality for one cone within the first Brillouin
zone and ``Dresselhaus-like'' chirality for the other one. As a
result, the net pseudospin accumulation {\it vanishes} when
contributions from the two cones are added together, cf.
Eq.~(\ref{result}).

The situation changes dramatically for topological insulators.
First, the chirality originates from {\it true} spin, accessible by
standard experimental techniques. Second, there is an {\it odd}
number of Fermi valleys per (geometric) surface. This is in a sharp
contrast to graphene which has two Dirac cones.  Note that despite
the equal number of ``species'' with opposite chiralities residing
on the opposite surface, one can probe them {\it individually}, by
virtue of their {\it spatial} separation.

\begin{figure}
\centerline{\psfig{file=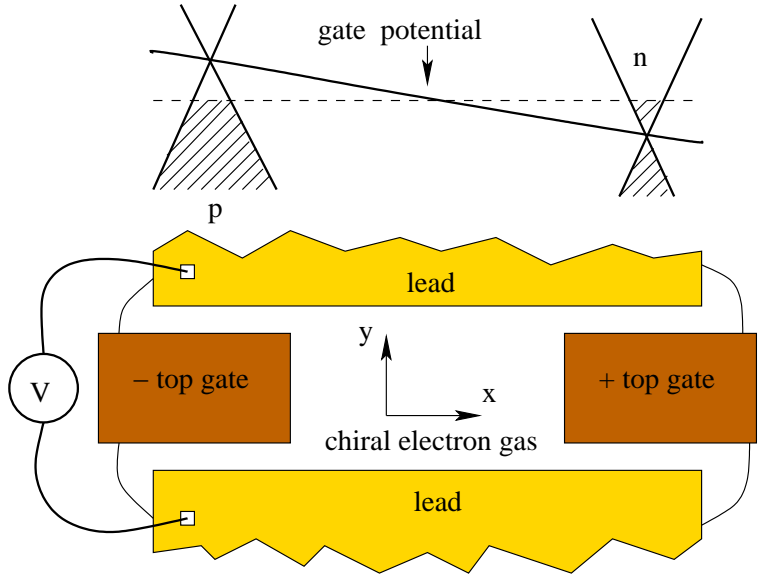,width=10.cm}}
 \caption{ Proposed experimental setup for creating a $n-p$ junction
on the surface of topological insulator. Voltage $V$ applied
between the leads drives electric current in $y$-direction and
creates a finite spin density along the junction (along the line
$x=0$).} \label{fig1.9}
\end{figure}

Since the Fermi energy in topological insulators lies within the
bandgap for 3D bulk electronic states, electron transport occurs
only across the crystal surface. While not yet implemented in
practice to our knowledge, it makes possible local control (via
 gate electrodes) of spatial distribution of electron
density. We envisage the following experimental setup, see
Fig.~\ref{fig1.9}. Metallic contacts are attached to the surface of
topological insulator and drive dc electric current along
$y$-direction. Gates are positioned (without direct contact) some
distance above the surface. Application of electric potentials to
the gates modulates position of electronic bands. Here, similarly to
the rest of the paper, we assume that the topological insulator is
disorder-free (ballistic). At the present time it is unclear how
well this assumption is satisfied in contemporary samples, but
undoubtedly sample quality is only going to improve in the near
future.

Effective low-energy Hamiltonian for Bi$_2$Te$_3$~\cite{Fu}
(neglecting cubic terms)
 \bq\label{Hinsul}
H=-iv(\sigma_x\partial_y -\sigma_y\partial_x) +U(x),
 \ee
 has a single Dirac point whose elevation is determined by the gate
 potential $U(x)$. There are two cases of interest here: \\
 i) Potential $U(x)$ is not very strong, so that the Dirac point
 always lies below the Fermi-energy (which is also normal situation
 for
 ungated material). \\
 ii) Potential is strong enough to lift the Dirac point {\it above}
 the Fermi level in some region on the surface, thus forming a
 $p$-$n$ junction.

We now present our general method before analyzing these two cases
separately. The method extends the approach of Sec.~\ref{Hard
Wall} for the Hamiltonian (\ref{Hinsul})
and we present it here in detailed form for the reader's
convenience. Utilizing the fact that in a ballistic system $k_y$
is the integral of motion, we write the electron operators in the
mixed representation,
 \bq
\hat{\psi}(\vec{r})=\sum_{k_y}\hat{c}_{k_y}(x)e^{ik_y y}.
 \ee
Similarly, spin density in the mixed representation reads
 \bq
 \label{topo_spin}
\vec{s}(k_y,x)=\langle \hat{c}_{k_y}^\dagger(x)
\vec{\sigma}\hat{c}_{k_y}(x)\rangle
 \ee
and satisfies the equation
 \bq
\partial_t s_y(k_y,x)=-\partial_x J_x^y(k_y,x) -2vk_ys_z(k_y,x).
 \ee
Here $J_x^y$ is spin current operator (in general
$J_j^i=\fr{1}{2}(\hat{\sigma}_i \hat{v}_j+
\hat{v}_j\hat{\sigma}_i)$),
 \bq
J_x^y(k_y,x)=-v\langle \hat{c}_{k_y}^\dagger(x)
\hat{c}_{k_y}(x)\rangle.
 \ee
In a steady state the left-hand side of Eq.~(\ref{topo_spin})
vanishes and we arrive at
 \bq
s_z(k_y,x)=-\fr{1}{2vk_y}\partial_x J_x^y(k_y,x).
 \ee
This yields the net spin polarization across the profile of the gate
potential $U(x)$
 \bq
 \label{topo_net}
\int_{-\infty}^\infty s_z(x) dx=-
\fr{1}{2v}\sum_{k_y}\fr{1}{k_y}[J_x^y(k_y,\infty)
-J_x^y(k_y,-\infty)],
 \ee
expressing it via the values of spin current far away from the
region where $U(x)$ is varied,
 \bq
 \label{topo-current}
J_x^y(k_y,\pm \infty)=-v\sum_{k_x} n(k_x,k_y,\pm \infty),
 \ee
with $n(k_x,k_y,\infty)$ denoting the corresponding distributions of
electrons.

The summation in Eqs.~(\ref{topo_net})-(\ref{topo-current}) has to
be performed over ``uncompensated'' states that originate from the
left lead and belong to the energy interval $E_F<E<E_F+eV$, where
$V$ is the bias applied between electrical contacts. (The
uncompensated states originate in the contact that has higher
chemical potential and propagate towards the other contact.) The
subsequent analysis will be performed separately for the two cases
mentioned above.

i) $n$-$n$ junction. When gate potential $U(x)$ is weaker than
needed to elevate Dirac point above the Fermi-energy the electric
current is carried by electrons only. Still, a smooth step-like
potential $U(x)$ create a ``junction'' between half planes with
different values of the Fermi momenta, $k_L$ and $k_R$,
respectively.

Since $ dk_x={dE k}/{k_x v}$ and $dE \to eV$ the summation over
$k_x$ is performed according to
 \bq
\sum_{k_x}\rightarrow \fr{eVk_n}{\pi\sqrt{k_n^2-k_y^2}},
 \ee
(note that both $k_x>0$ and $k_x<0$ contribute to this expression).
From Eq.~(\ref{topo_net}),
 \bq\label{insulatorI}
\int_{-\infty}^\infty s_z(x) dx=\fr{eV}{2\pi v}\left\{
\int_0^{k_L}\fr{dk_y}{k_y}\fr{k_L}{\sqrt{k_L^2-k_y^2}} -
\int_0^{k_R}\fr{dk_y}{k_y}\fr{k_R}{\sqrt{k_R^2-k_y^2}}\right\}.
 \ee
Regularizing these formally divergent integrals (by assuming the
same lower cutoff which subsequently drops out from the difference
of the two terms) we find the amount of net spin accumulated at the
junction,
 \bq\label{insulatorV}
\int_{-\infty}^\infty s_z(x) dx=\fr{eV}{2\pi v}\ln\left(
\fr{k_L}{k_R}\right).
 \ee
In addition, if the gate potential changes smoothly on the scale
of $1/k_F$ the local spin polarization can be written as, \bq
s_z(x) =\fr{eV ~\partial_x U(x)}{2\pi v^2 k_F(x)} . \ee

ii) $p$-$n$ junction. In the real experiment the strength of the
gate potential $U(x)$ can be made significant enough to  lift
Dirac point above the Fermi energy over some region of the
surface.
For this setup we find
 \bq\label{insulatorII}
\int_{-\infty}^\infty s_z(x) dx=\fr{eV}{2\pi v}\left\{
\int_0^{k_n}\fr{dk_y}{k_y}\fr{k_n}{\sqrt{k_n^2-k_y^2}} -
\int_{-k_p}^{0}\fr{dk_y}{k_y}\fr{k_p}{\sqrt{k_p^2-k_y^2}}\right\}
.
 \ee
This result differs from Eq.~(\ref{insulatorI}), and from the spin
accumulation around $v_F(x)=\lambda$ considered in
section~\ref{Soft Wall}, in that here the excess electrons to the
left and to the right of the potential step have similar velocity,
$v_y$, but opposite momentum, $p_y$, in the direction of current.
In Eq.~(\ref{insulatorII}) the excess electrons originate from the
lead $E_F+eV$ and thus have to have $v_y>0$. On the contrary, in
the case of edge spin accumulation (Sec.~\ref{Soft Wall}) the
electrons arrive to the strip $0<v_F(x)<\lambda$ from the bulk
2DEG, as is shown in Fig.~\ref{fig1.8}, and thus must carry the
bulk momentum $p_y>0$, but may have the "wrong" sign of velocity.
As a result the spin density in Eq.~(\ref{insulatorII}) has the
same sign on both sides of the $n-p$ junction and two divergent
logarithms add.
Introducing a proper quantum mechanical cutoff to these integrals
yields
 \bq\label{insulatorVV}
\int_{-\infty}^\infty s_z(x) dx=\fr{eV}{2\pi v}\ln\left( \fr{4k_p
k_n}{q_{\rm{min}}^2}\right) \ , \ q_{\rm{min}}^2=\fr{1}{\hbar
v}\left. \fr{dU}{dx}\right|_{x=0} .
 \ee
Electrons with very small $k_y$ start to tunnel thorough the $n-p$
junction, which would change our semiclassical predictions. The
rude estimate of the quantum mechanical tunnelling exponent gives
$\kappa =\int_0^{k_y} k_x(x)dx =k_y^2\hbar v/(2dU/dx)$. Requiring
a small tunnelling probability, $e^{-\kappa}\ll 1$, gives the
above cutoff $q_{\rm{min}}$.

We may now compare the strength of the Spin-Hall effect in
semiconductor heterostructures Eq.~(\ref{result}) with the result
for topological insulators
Eqs.~(\ref{insulatorV},\ref{insulatorVV}). The main difference is
that in conventional semiconductors the chiral electron states
appear because of the weak spin-orbit interaction. Consequently
the result Eq.~(\ref{result}) acquires a small factor
$(\lambda^2-\lambda_D^2)/v_F^2$. There is now such suppression for
the surface states in topological insulators\footnote{Spin
accumulation Eq.~(\ref{result}) and
Eqs.~(\ref{insulatorV},\ref{insulatorVV}) depends also on the
value of the Fermi velocity. The value of $v_F$ in
Eq.~(\ref{result}) depends on the bulk density of 2DEG. In real
experiment however, the Fermi velocity in our two examples will
probably be of the same order of magnitude.}.

\section{Summary}

In this paper we discussed several nonequilibrium spin-related
phenomena occurring in 2-dimensional chiral electron systems.

We were mostly interested in the effects which may be explained in
terms of semiclassical electron motion in smooth potentials. In
this case it is easy (at least theoretically) to produce the
strongly in-plane spin polarized currents
(Sec.~\ref{Semiclassics}).

The main part of the paper was devoted to calculation of the out
of plane spin density $\langle\sigma_z\rangle$. It is the out of
plane spin component, that is usually investigated in the
experiment via the measuring of the optical Kerr
rotation\cite{exp1}.

Finally we discussed the mesoscopic spin-Hall effect in novel
materials such as graphene and topological insulators.
Experimental investigation of pseudospin-Hall effect might prove
not to be easy. On the contrary, observation of spin-Hall
$\sigma_z$ accumulation at the surface of 3D topological insulator
looks most appealing. While the net spin accumulation in
semiconductor heterostructures is of the second order in small
spin-orbit interaction, $s\sim\lambda^2/v_F^3$ (\ref{result}),
there is no such restriction for topological insulator, $s\sim
1/v$ (\ref{insulatorV},\ref{insulatorVV}). As a result, the spin
accumulation in the latter case should exceed that possible in
traditional 2-dimensional electron gas by orders of magnitude.

\section*{Acknowledgments}
We gratefully acknowledge our collaborator V.~Zyuzin, who
participated in obtaining a number of previously published results
reviewed in this paper; we also have benefited from discussions
with A.~Andreev, G.E.W.~Bauer, C.W.J.~Beenakker, B.I.~Halperin,
M.~Raikh and O.~Starykh. P.G.S. was supported by the SFB TR 12;
E.G.M. was supported by the DOE, Office of Basic Energy Sciences,
Award No.~DEFG02-06ER46313.

\bibliographystyle{ws-rv-van}
\bibliography{ws-rv-sample}

\printindex                         

\end{document}